\documentclass{aa}
\usepackage{graphicx}
\usepackage{txfonts}

\begin{document}

\title{Feedback from Intra-Cluster Supernovae on the ICM in 
Cooling Flow Galaxy Clusters}

\author{Wilfried Domainko
 \and Myriam Gitti
 \and Sabine Schindler
 \and Wolfgang Kapferer}

\institute{Institut f\"ur Astrophysik, Leopold-Franzens Universit\"at Innsbruck
Technikerstra\ss e 25, A-6020 Innsbruck, Austria}

\offprints{Wilfried Domainko,
\email{wilfried.domainko@uibk.ac.at}}

\authorrunning{Domainko et al.}
\titlerunning{Feedback from Intra--cluster Supernovae on the ICM in Cooling 
Flow Galaxy Clusters}
\date{Received / Accepted}
\abstract{
We study the effect of heating and metal enrichment from supernovae (SNe) 
residing between galaxies on the Intra-Cluster Medium (ICM). 
Recent observations indicate that a considerable fraction 
($\sim$20\%) of the SN Ia parent stellar population in galaxy clusters 
is intergalactic.
By considering their effect on the relaxed progenitors of cooling flow clusters
we propose that intra--cluster SNe can act as a distributed heating source 
which may influence the initial stages of the formation of cooling flows.
We investigate the increase in cooling time as a function of the
energy input supplied by SNe and their assumed spatial distribution, and
conclude that intra--cluster SNe represent a heating source which 
in some clusters can cause a delay of the formation of cooling flows. 
This would imply that some cooling flows are younger than previously 
thought. We also discuss the impact that a large population of intra--cluster 
SNe could have on the chemical evolution of the ICM in cooling flow clusters.
}
\maketitle


\section{Introduction}

In the central regions of clusters of galaxies the gas density is often high
enough that the radiative cooling time due to X--ray emission is shorter
than the Hubble time. 
Therefore, without any balancing heating mechanisms, the gas should cool 
down and flow slowly inwards in order to maintain the hydrostatic equilibrium.
Recent X--ray observations with {\it Chandra} and {\it XMM--Newton} 
(e.g. David et al. \cite{david}, Johnston et al. \cite{johnstone}, 
Peterson et al. \cite{peterson03} and references therein)
have altered this simple picture of a steady cooling flow (for a review of the 
standard cooling flow model see Fabian \cite{fabian94}).
The lack of observations of cooler gas represent an open question which is 
often referred to as the so called {\it 'cooling flow problem'}. 
This topic is still lively debated and has recently triggered the development 
of a variety of theoretical models.
 
There are two main approaches to solve this problem. 
The first solution is that the gas does cool, but we do not observe it. 
Different possibilities have been investigated in this context, including
absorption (Peterson et al. \cite{peterson01}, Fabian et al. \cite{fabian01}),
inhomogeneous metallicity (Fabian et al. \cite{fabian01}, Morris \& Fabian
\cite{morris03}), and the emerging of the missing X--ray luminosity
in other bands, like ultraviolet, optical and infrared (due to
mixing with cooler gas/dust, Fabian et al. \cite{fabian01}, \cite{fabian02-a},
Mathews \& Brighenti \cite{mathews03}), and radio (due particle 
re--acceleration, Gitti et al. \cite{gitti02}, \cite{gitti04}).
The second solution is that an additional heating mechanism
which balances the cooling is acting in the intra--cluster medium (ICM). 
Proposed heating mechanisms include heating through processes associated 
with relativistic AGN outflows (e.g., Rosner \& Tucker \cite{rosner}; 
Tabor \& Binney \cite{tabor}; Churazov et al. \cite{churazov}; 
Br{\"u}ggen \& Kaiser \cite{brueggen}; Kaiser \& Binney \cite{kaiser};  
Ruszkowski \& Begelman \cite{ruszkowski}; Brighenti \& Mathews 
\cite{brighenti}), 
electron thermal conduction from the outer regions of clusters
(Tucker \& Rosner \cite{tucker}; Voigt et al. \cite{voigt}; 
Fabian et al. \cite{fabian02-b}; Zakamska \& Narayan \cite{zakamska}),
continuous subcluster merging (Markevitch et al. \cite{markevitch}),
and contribution of the gravitational potential of the cluster core 
(Fabian \cite{fabian03}).
In general, proposed heating sources face problems with the spatial 
distribution of the heating rate necessary to balance the cooling, 
which is a crucial requirement for a successful model.
The further possibility of combining these two approaches has 
been investigated in the context of the moderate cooling flow model
(Soker \cite{soker01}, \cite{soker04}), in which  
the heating does not completely balance 
cooling and the gas cools at lower rates, leading to a better agreement 
with the recent observations.

In this letter we discuss a novel approach in the context of the 
heating scenario --- we consider intra--cluster supernovae as an 
energy source to delay the formation of cooling flows in relaxed clusters. 
The effect of heating by supernovae (SNe) in cooling flows has already 
been studied in previous works (e.g., Brighenti \& Mathews \cite{brighenti},
McNamara et al. \cite{mcnamara}), although the 
contribution of intra--cluster SNe has never been included.
The existence of such a population of supernovae (SNe) is suggested by 
recent observations showing evidence of an intra--cluster 
stellar population (ICSP) in nearby clusters, which is also predicted by
N--body simulations (e.g. Napolitano et al. \cite{napolitano} and references
therein).
Feldmeier et al. (\cite{feldmeier}) found many 
intra--cluster planetary nebulae in 
the Virgo cluster while Theuns \& Warren (\cite{theuns})  
found the same for the Fornax cluster. 
Arnaboldi et al. (\cite{arnaboldi}) used planetary nebulae in the Virgo 
cluster as a tracer for the diffuse intra--cluster light and determined 
limits for the total luminosity of the ICSP. 
They find an amount of 10-40\%
of the total stellar population. 
Furthermore, a population of red giant stars in the Virgo cluster
was discovered with HST observations (Ferguson et al.\cite{ferguson})
and even direct evidence for intra--cluster SNe was found (Gal-Yam et al.  
\cite{gal-yam}).
In general, these observations indicate that the 
ICSP consists of an old stellar population, therefore only SNe type 
Ia should occur there due to their long evolution time. \\
$H_0 = 50 \mbox{ km s}^{-1} \mbox{ Mpc}^{-1}$ is used throughout the 
paper.


\section{Feedback from intra--cluster SNe}

\subsection{Heating Rate}

As discussed in the Introduction, our approach to the
cooling flow problem is to consider  an energy source to delay the 
evolution of relaxed clusters (hereafter
progenitor clusters) to  cooling flow clusters. 
In the ideal case, a heating source which is able to supply the
energy lost to X--rays on the same timescale of the emission process
would completely prevent the formation of cooling flows. 
Note that this approach is different from a scenario stopping cooling 
flows after they have long been established.
Here we suggest that intra--cluster SNe  
may contribute to that heating source. Indeed, their rate is not 
triggered by cooling flows themselves and therefore,
contrarily to other heating sources (e.g.  
AGNs), 
intra--cluster SNe can act also in the progenitor clusters.
Four main reasons support our general idea:
1)
new observations show that a considerable fraction ($ \sim$ 20\%) of the SN Ia 
parent stellar population in clusters is intergalactic (Gal-Yam et al.  
\cite{gal-yam});
2)
SNe exploding in the ICM can heat the ambient medium very efficiently 
and their metal rich material will be ejected directly into the ICM;
3)
intra--cluster SNe provide a distributed heating source;
4)
SN heating can act over a long period of time.

To investigate whether intra--cluster SNe can have a significant
impact on the evolution of the progenitor clusters, we compare the energy lost 
by these clusters in a Hubble time with the energetics of SN heating.
One way to estimate the X--ray emission of the progenitor clusters is to start
from observations of cooling flow clusters and extrapolate the fit of the 
surface brightness in the outer regions (regions out of the cooling radius are 
unaffected by the cooling flow) to the center. 
Note that this results in a flat density distribution in the central 
region of the cluster. 
In particular, by adopting the density distribution  
derived in this way and the cluster temperature observed outside the 
cooling region, we determine the X--ray luminosity up to 
a certain radius by integrating over the X--ray emissivity. 
X--ray luminosities are multiplied by $10^{10}$ years to compare 
them with the SN heating provided in a Hubble time. 
To get an impression of the importance of this effect we give here two 
examples. 
Values are derived in the central 130 kpc. Input parameters 
are in brackets. 
1. Perseus cluster ($ T=6.2$ keV, core radius: $r_{\rm c} = 280$ kpc, 
$\rho(0) = 
4.05 \times 10^{-3} \mbox{cm}^{-3}$, $\beta=0.87$; Churazov et al. 
\cite{churazov03}):
$L_{\rm X} \sim 6.8 \times 10^{43}$ erg/s or $\sim 2.1 \times 10^{61}$ erg
 in $10^{10}$ years. 
2. Centaurus cluster ($ T=3.4$ keV, $r_{\rm c} = 110$ kpc, $\rho(0) = 2 \times 
10^{-3} \mbox{cm}^{-3}$, $\beta=0.66$; values were derived from 
Allen \& Fabian \cite{allen94} and Sanders \& Fabian \cite{sanders}): 
$L_{\rm X} \sim 1.4 \times 10^{43}$ erg/s 
or $\sim 4.4 \times 10^{60}$ erg in $10^{10}$ years. 

The energetics of SN heating can be estimated
from the central abundance peak observed in cooling flow clusters.
Various authors argued that most of the central  
iron abundance is produced by SN Ia (e.g. B{\"o}hringer et al. 
\cite{boringer}) therefore we calculate the heating due to SN Ia
by assuming an iron production of $0.7 M_{\sun}$ per SN 
(e.g. Renzini et al. \cite{renzini}).
As most of the energy from SNe exploding into the thin hot ICM will be 
transformed into thermal energy, a typical SN Ia will deposit a 
total energy of the order of $\sim 10^{51}$ erg in the ICM 
(a detailed discussion on SN remnants in hot thin 
environments can be found in Dorfi \& V{\"o}lk \cite{dorfi}).
On this basis Sasaki (\cite{sasaki}) already proposed 
intra--cluster SNe as a possible extra energy source in clusters.
By adopting the iron excess found by B\"ohringer et al. 
(\cite{boringer}) in the central 130  kpc
\footnote{Quantities were scaled accordingly to the value of $H_0$ adopted}
we derive the number of SNe necessary to produce it and then, by knowing 
the energy injected per SN, we calculate the total energy provided.
For the two examples we find: 1. Perseus cluster:$\sim 2.4 \times 10^{9}  
M_{\sun}$ Fe, $\sim 3.4 \times 10^{9} $ SNe Ia leading to a total heating of
$ \sim 3.4 \times 10^{60} $ erg. 2. Centaurus cluster:
$\sim 1.4 \times 10^{9}  
M_{\sun}$ Fe, $\sim 2 \times 10^{9}  $ SNe Ia leading to a total heating of
$ \sim 2 \times 10^{60} $ erg.
This corresponds to $\sim16\%$ (Perseus) and $\sim45\%$ (Centaurus)
of the energy emitted in X--rays by the progenitor clusters.
So, while in Perseus the effect is not so important, in the case of Centaurus
the additional heating by SNe may be significant. In general, the importance
of the effect will depend on the particular cluster.

Note that the complete compensation of the energy loss would be in disagreement
with the observations which show that the ICM -- at least at some level --
does cool.

In the previous estimate we did not discriminate between the iron
contribution by galactic and intra--cluster SNe.
Even though the distinction between the remote
envelopes of the cD galaxy and the ICSP may simply be a matter of semantics
(Gal-Yam et al. \cite{gal-yam}), we stress that the fraction of iron 
produced by intra--cluster SNe will be significantly higher than 20\%, 
as a substantial part of the iron produced by galactic SNe 
will remain in their host galaxies. 
SNe explosions associated with the cD galaxy stellar population 
will also contribute to that heating (McNamara et al. 
\cite{mcnamara}, Wise et al. \cite{wise}). 
The SN rate for SN Ia might be even higher in the past 
by a factor of 10 (Renzini et al. \cite{renzini}) which will also have an 
important impact on the time dependence of the problem.
       
In our model we consider a single phase ICM of the progenitor cluster 
so the X--ray emission, which is $\propto \rho^2$, 
is mainly a function of the distance $r$.
The density distribution $\rho(r)$ is described by a $\beta$ profile 
(Cavaliere \& Fusco-Femiano  \cite{cavaliere}).
Different distributions of SN heating rate are described with King profiles 
with variable exponent  which 
can be written in a similar way as the 
$\beta$ model:
\begin{equation}
H_\mathrm{SN}(r)= \frac{H_\mathrm{SN}(0)}{[1+(r/r_c)^2]^{3\beta_\mathrm{SN}/2}}
\label{h.sn}
\end{equation}
Here $H_\mathrm{SN}(r)$ is the radial dependence of the SN heating rate, 
$r_\mathrm{c}$
is the core radius and $\beta_\mathrm{SN}$ controls the outer 
slope of this distribution.

In order to investigate the effect of intra--cluster SNe on cooling flows,
we introduce the cooling time $t_\mathrm{cool,SN}$ which takes into account
the heating contribution from these sources:  
\begin{equation}
t_\mathrm{cool,SN}(r) = \frac{E_\mathrm{th}(r)}
{L_\mathrm{X}(r)-H_\mathrm{SN}(r)}
\end{equation}
where $E_\mathrm{th}$ is the thermal energy  of the ICM, 
$L_\mathrm{X}$ is the X--ray luminosity and 
$H_\mathrm{SN}$ is that of Eq. \ref{h.sn}.
  
\begin{figure}[ht]
\includegraphics{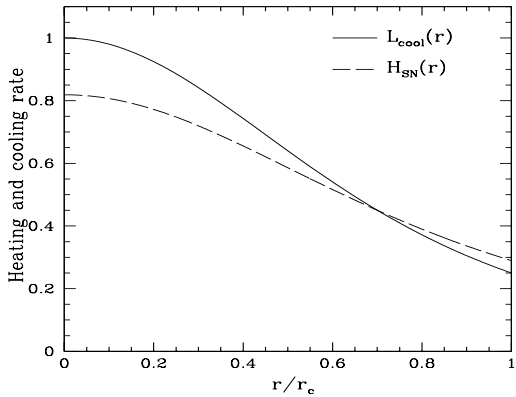}
\vspace{5cm}
\caption{An example of the spatial distribution for intra--cluster
SN heating rate and cooling rate due to X--ray emission. 
Both profiles are normalized to the same 
total energy when integrated over the core region of the cluster. 
The units on the y-axis are in arbitrary linear scale. 
For the cooling profile a $\beta$ model for the density with a 
$\beta_\mathrm{\rho}$ of 0.66 is chosen, while
for the SNe distribution a King profile with $\beta_\mathrm{SN}=1$ is used.}
\label{figure1}
\end{figure}

\begin{figure}[ht]
\includegraphics{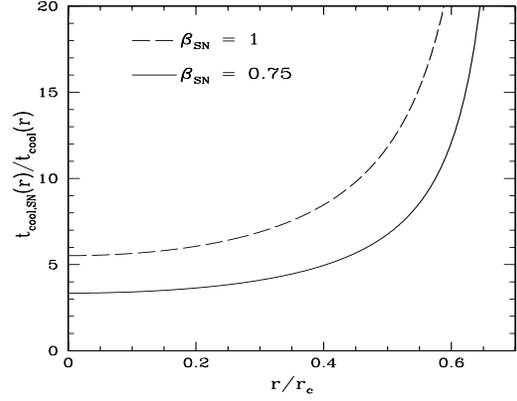}
\vspace{5cm}
\caption{The radial dependence of the introduced cooling time in units of 
the cooling time $t_{\rm cool}$. The two curves 
correspond to different $\beta_\mathrm{SN}$ for two different SN 
distributions.}
\label{figure2}
\end{figure}

Because of its dependency on the square of the density, the X--ray emission 
will be more peaked towards the cluster center with respect 
to the SN heating rate. 
By normalizing such that the total energy input due 
to SN heating (integrated over the cluster core region) balances 
the total X--ray luminosity (as an ideal case where heating 
balances cooling), 
this means that for any realistic density and SNe distribution 
the energy loss due to cooling is more efficient than the SN heating rate 
near the cluster center whereas the SN heating rate dominates at bigger radii 
(see Fig. \ref{figure1}).
~\\
The radial dependence of our defined cooling time is shown in Fig. 
\ref{figure2}, where we plot $t_\mathrm{cool,SN}(r)$
as a fraction of the cooling time $t_{\rm cool}(r)$ calculated without the 
heating contribution.

\begin{figure}[ht]
\includegraphics{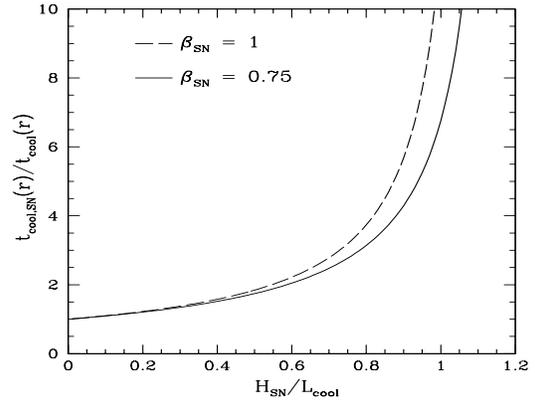}
\vspace{5cm}
\caption{The delay of the cooling process relative to an undisturbed 
formation of a cooling flow. 
On the x axis the fraction of total heating energy in units of the 
total energy loss due to cooling is shown. 
The two curves correspond to different $\beta_\mathrm{SN}$
for the SN distributions.
Results are given for a radius of 0.5 core radii. For the density 
distribution $\beta_\mathrm{\rho}= 0.66$ is chosen.}
\label{figure3}
\end{figure}

Under these assumptions we find that the SN heating rate will not 
prevent the cluster center from cooling, but it will 
lengthen the timescale on which the gas cools.
The importance of this effect can vary 
from cluster to cluster, depending on 
the energy input supplied by SNe and their assumed spatial distribution.
To show this,
we plot the increase in cooling time for various parameters 
(Fig.\ref{figure3}). 
For example, by assuming $\beta_\mathrm{SN}=1$
a total SNe heating rate corresponding to a fraction  $\sim 0.6$ of the
total X--ray luminosity would produce a factor of 2 increase in 
cooling time, which can be very important for the outers parts of a cluster 
core (Fabian et al. \cite{fabian03}).

As a result of these considerations, we conclude that intra--cluster 
SNe can provide an extended heating input into the ICM.
The significance of this additional heating depends on the particular
cluster, as we showed in two different examples, and in some cases it may cause
a delay of the cooling process. 
This would imply that some cooling flows are younger than previously 
thought, as already argued by several authors (e.g. Allen et al. \cite{allen}, 
Johnston et al. \cite{johnstone}).

\subsection{Iron enrichment}

In Sect. 2.1 we estimated the SN heating from the observed iron 
masses in the central regions of cooling flows. 
We point out that the observed iron masses might be lower limits
to the total amount of iron produced, because some of the iron might be 
deposited by fueling the central AGN and star formation (McNamara et al.
\cite{mcnamara}). This would mean that the energy injected might be higher
than estimated from the iron mass.

Another consequence of the presence of a large population of intra--cluster 
SNe could be that of producing a metallicity gradient in the ICM of cooling
flow clusters, which is an alternative view to the scenario where
the central abundance peak originates from the stellar population of the cD
galaxy (B{\"o}hringer \cite{boringer}).  
Indeed, a gas flow passing SN explosions will be loaded with
iron thus resulting in a central abundance peak when moving towards the cluster
center. Different distributions of SNe and gas density will
lead to a similar effect. We stress that metallicity gradients are in fact
observed in cooling flow clusters (e.g., De Grandi \& Molendi 
\cite{degrandi}), even though a more detailed model 
with various enrichment processes (ram-pressure stripping, 
galactic winds, etc.) would be required in order to compare directly the 
predicted metallicity gradient with that observed.

Finally, we note that
material from SNe exploding in between the galaxies will 
eject the iron directly into the ICM. 
Therefore, the remnants of intra--cluster SNe may end up as highly enriched 
bubbles in the ICM and this could result in rapid cooling of the highly 
enriched material (Fabian et al. \cite{fabian01}, 
Morris \& Fabian \cite{morris03}).

\section{Conclusions}

In this paper we investigate the feedback from intra--cluster SNe on 
the ICM in cooling flow clusters.
By considering their effect on the relaxed progenitors of cooling flow clusters
we suggest that intra--cluster SNe represent a heating source 
for the ICM:
the significance of this additional heating depends on the particular
cluster and in some cases it may cause a delay of the formation of cooling 
flows.
In general the proposed scenario does not solve the cooling flow
problem, so another heating mechanism is needed (e.g. AGNs).
However, since intra--cluster SNe can act as a distributed heating 
source once the cooling flow has started, they may represent an interesting
complement to heating models which consider central AGN activity.
Intra--cluster SNe might also have a remarkable impact on the chemical 
evolution of galaxy clusters and explain the presence of
metallicity gradients in cooling flows.
Recent observations of a population of intra--cluster SNe (Gal-Yam et al. 
\cite{gal-yam}) and the high efficiency of SN heating rate in an 
ambient hot, thin medium support our general idea. 
Numerical simulations show that the ICSP originates from tidal debris 
caused by galaxy-galaxy and galaxy-cluster interaction (Napolitano et al.
\cite{napolitano}).
One important open question in connection with the model is: will 
tidal interaction between galaxies disrupt or enhance the SN Ia
formation mechanisms? 
Gal-Yam et al.(\cite{gal-yam}) conclude
that the intra--cluster SN Ia rate should 
be comparable to the SN rate found in galaxies. 
The critical points for our model are the spatial distribution of the
SNe and the SN rate, which both strongly influence the heating scenario.
More observations of the ICSP and a more detailed modelling
are required in order to reach
a better understanding of this problem.

\begin{acknowledgements}
The authors thank the referee N. Soker for his 
very helpful comments that improved the paper.
The authors would also like to thank C.L. Sarazin for useful
discussions on cooling flows.
W.D. would like to thank E.A. Dorfi, A. Egger, E. v.Kampen, 
W. Kausch, S. Kimeswenger and S. Kreidl for helpful discussions.
M.G. would like to thank S. Dall'Osso and F. Brighenti
for many stimulating discussions and insightful comments.
This work was supported by the Austrian Science Foundation FWF under 
grant P15868.
\end{acknowledgements}

\end{document}